\documentclass[sigconf, screen]{acmart}

\acmSubmissionID{1012}

\usepackage{hyperref}
\usepackage{latexsym}
\usepackage{mathrsfs}
\usepackage{multirow}
\usepackage{amsmath}
\usepackage{amsfonts}
\usepackage{subfigure}
\usepackage{graphicx}
\usepackage{tabularx}
\usepackage[boxed,ruled,lined]{algorithm2e}
\usepackage{algorithmic}
\usepackage{comment}
\usepackage{balance}
\usepackage{bm}
\usepackage[inline]{enumitem}
\usepackage[skip=1mm]{caption}

\usepackage{acronym}
\usepackage{makecell}
\usepackage{listings}
\usepackage[most]{tcolorbox}

\usepackage{xcolor}

\lstdefinestyle{shell}{
    language=bash,        
    basicstyle=\ttfamily, 
    keywordstyle=\color{black},   
    stringstyle=\color{blue},   
    commentstyle=\color{gray},   
    backgroundcolor=\color{black!5}, 
    breaklines=true,      
    frame=single,         
    showstringspaces=false, 
    morekeywords={python},
}

\lstdefinelanguage{json}{
  basicstyle=\ttfamily\small,
  numbers=left,
  numberstyle=\tiny\color{gray},
  stepnumber=1,
  numbersep=5pt,
  showstringspaces=false,
  breaklines=true,
  literate=
   *{0}{{{\color{blue}0}}}{1}
    {1}{{{\color{blue}1}}}{1}
    {2}{{{\color{blue}2}}}{1}
    {3}{{{\color{blue}3}}}{1}
    {4}{{{\color{blue}4}}}{1}
    {5}{{{\color{blue}5}}}{1}
    {6}{{{\color{blue}6}}}{1}
    {7}{{{\color{blue}7}}}{1}
    {8}{{{\color{blue}8}}}{1}
    {9}{{{\color{blue}9}}}{1}
    {:}{{{\color{red}:}}}{1}
    {,}{{{\color{red},}}}{1}
    {"}{{{\color{orange}"}}}{1},
}

\newtcolorbox{promptbox}{
  colback=blue!9!white,   
  colframe=blue!55!white, 
  coltitle=black,         
  fonttitle=\bfseries,    
  title=Prompt,           
  boxrule=0.8pt,          
  arc=3mm,                
  top=4pt, bottom=4pt, left=6pt, right=6pt, 
}

\newcommand{\eg}{\emph{e.g., }}

\newcommand{\wrt}{\emph{w.r.t. }}

\acrodef{IR}{information retrieval}

\usepackage{amsmath,amsfonts,bm}

\def\eqref#1{equation~\ref{#1}}

\def\1{\bm{1}}

\DeclareMathAlphabet{\mathsfit}{\encodingdefault}{\sfdefault}{m}{sl}
\SetMathAlphabet{\mathsfit}{bold}{\encodingdefault}{\sfdefault}{bx}{n}

\DeclareMathOperator*{\argmin}{arg\,min}

\author{Chen Xu}
\orcid{0000-0002-3070-9358}
\affiliation{%
  \institution{Renmin University of China}
  \city{Beijing}
  \country{China}
}
\email{xc\_chen@ruc.edu.cn}

\author{Zhipeng Yi}
\orcid{0009-0009-4049-9540}
\affiliation{
  \institution{University of Science and Technology of China}
  \city{Hefei}
  \country{China}
}
\email{zhipengyi@mail.ustc.edu.cn}

\author{Ruizi Wang}
\orcid{0009-0000-6945-4404}
\affiliation{
  \institution{University of Science and Technology of China}
  \city{Hefei}
  \country{China}
}
\email{wangrz_ustc@mail.ustc.edu.cn}

\author{Wenjie Wang}
\authornote{Jun Xu and Wenjie Wang are corresponding authors.}
\orcid{0000-0002-5199-1428}
\affiliation{
  \institution{University of Science and Technology of China}
  \city{Hefei}
  \country{China}
}
\email{wenjiewang96@gmail.com}

\author{Jun Xu}
\authornotemark[1]
\orcid{0000-0001-7170-111X}
\affiliation{
  \institution{Renmin University of China}
  \city{Beijing}
  \country{China}
}
\email{junxu@ruc.edu.cn}

\author{Maarten de Rijke}
\orcid{0000-0002-1086-0202}
\affiliation{
 \institution{University of Amsterdam}
 \city{Amsterdam}
 \country{The Netherlands}
}
\email{m.derijke@uva.nl}

\renewcommand{\shortauthors}{Chen Xu et al.}

\copyrightyear{2026}
\acmYear{2026}
\setcopyright{cc}
\setcctype{by}
\acmConference[WWW '26]{Proceedings of the ACM Web Conference 2026}{April 13--17, 2026}{Dubai, United Arab Emirates}
\acmBooktitle{Proceedings of the ACM Web Conference 2026 (WWW '26), April 13--17, 2026, Dubai, United Arab Emirates}
\acmPrice{}
\acmDOI{10.1145/3774904.3792216}
\acmISBN{979-8-4007-2307-0/2026/04}



\title{User Addiction in Short-Video Platforms: Unveiling Patterns and Simulating Behaviors}
\title{Unveiling User Addiction Behaviors in Short-Video Platforms}
\title{Unveiling and Simulating the Economic Mechanisms of User Addiction Behaviors in Short-Video Platforms}
\title[Unveiling and Simulating Short-Video Addiction Behaviors via Economic Addiction Theory]{Unveiling and Simulating Short-Video Addiction Behaviors\\ via Economic Addiction Theory}

\begin{document}

\begin{abstract}
Short-video applications have attracted substantial user traffic. 
However, these platforms also foster problematic usage patterns, commonly referred to as short-video addiction, which pose risks to both user health and the sustainable development of platforms. 
Prior studies on this issue have primarily relied on questionnaires or volunteer-based data collection, which are often limited by small sample sizes and population biases. 
In contrast, short-video platforms have large-scale behavioral data, offering a valuable foundation for analyzing addictive behaviors.
To examine addiction-aware behavior patterns, we combine economic addiction theory with users’ implicit behavior captured by recommendation systems. Our analysis shows that short-video addiction follows functional patterns similar to traditional forms of addictive behavior (e.g., substance abuse) and that its intensity is consistent with findings from previous social science studies. To develop a simulator that can learn and model these patterns, we introduce a novel training framework, AddictSim. To consider the personalized addiction patterns, AddictSim uses a mean-to-adapted strategy with group relative policy optimization training. Experiments on two large-scale datasets show that AddictSim consistently outperforms existing training strategies.
Our simulation results show that integrating diversity-aware algorithms can mitigate addictive behaviors well.

\end{abstract}

\maketitle

\section{Introduction}

In recent years, short-video applications, exemplified by platforms such as TikTok and Xiaohongshu, have witnessed a rapid surge in popularity across web ecosystems. While these platforms attract substantial user traffic through rapid content consumption, they also foster problematic usage patterns, often referred to as the \textit{short-video addiction} problem~\cite{ye2022effects, zhang2019exploring}. Such a problem refers to a compulsive tendency to continuously consume short videos, which impacts both users’ cognitive development, psychological health, and even the web ecosystem~\cite{ye2022effects, feng2024cognitive, chen2023effect}. However, the short-video addiction problem remains underexplored, highlighting the necessity for systematic research and intervention.

Previous studies on short-video addiction have predominantly relied on questionnaires, volunteer-based data collection, and self-report studies~\cite{ye2025short, zhang2019exploring, ye2022effects, feng2024cognitive}. Such methods are often limited by small sample sizes and potential biases arising from the characteristics of the target populations. In contrast, web-based short-video platforms generate large-scale behavioral data covering a broad spectrum of users, offering a more comprehensive perspective for analyzing addictive behaviors. To better analyze such behaviors, in this paper, we propose a two-step analysis paradigm:  (1)~analyzing behaviors with economic theory to define the objective, (2)~enabling a simulator to learn this objective.

\begin{figure}
    \centering
    \includegraphics[width=0.9\linewidth]{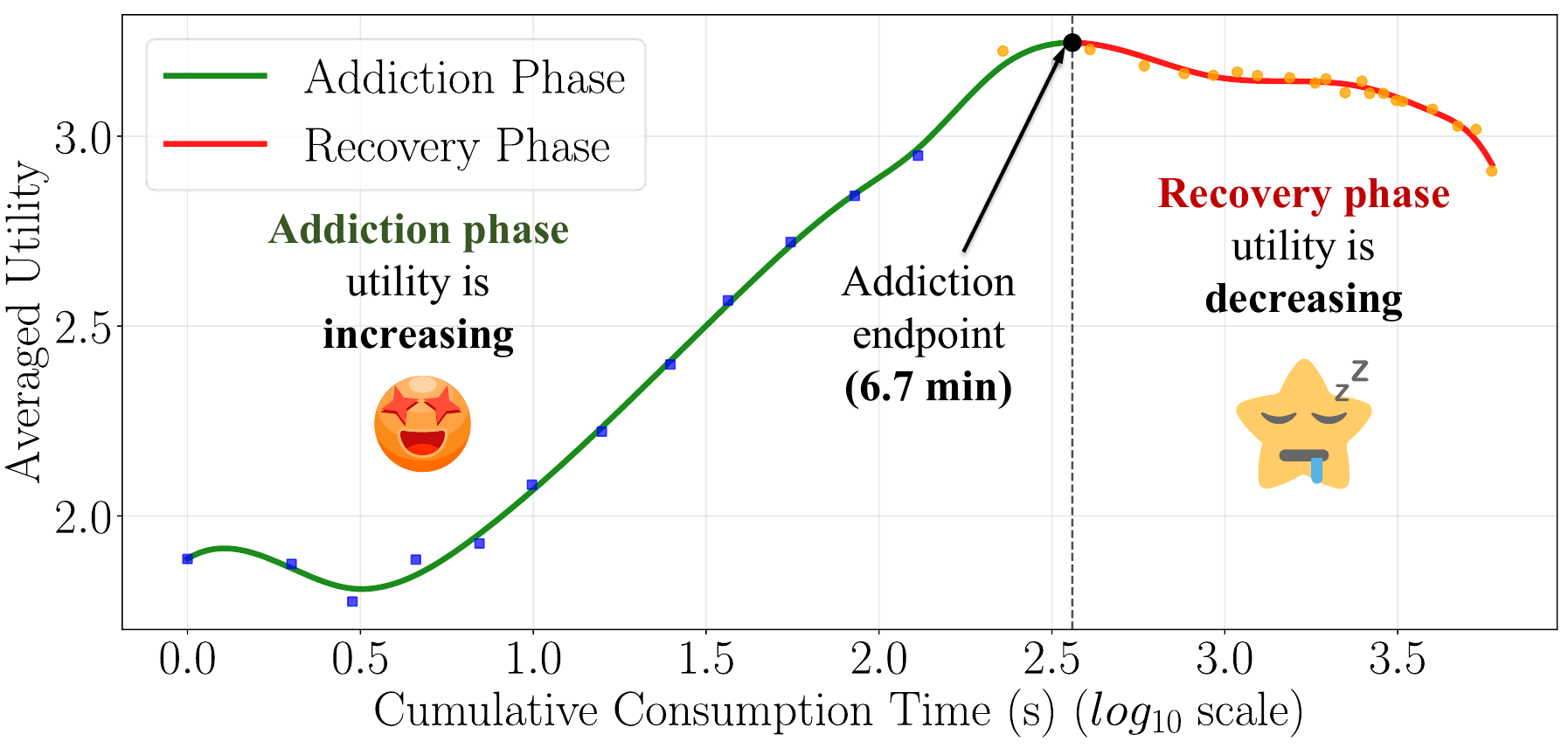}
    \caption{Addiction patterns on a popular short-video platform. The x-axis represents cumulative user watch time, and the y-axis represents user utility.}
    \label{fig:intro}
\end{figure}

\emph{Step 1: Addiction behavior discovering (Section~\ref{sec:addition}).} 
Firstly, addiction behaviors can be formalized based on the economic theory of addiction~\cite{west2013addictiontheory}, which posits that past consumption of a good increases current utility, thereby fostering habitual or compulsive behavior. However, directly applying this theory is challenging because real-world datasets in recommendation systems (RS) typically record only implicit interaction (\eg clicks and watch times) rather than explicit economic utilities.
To overcome this limitation, we observe that implicit feedback can still be interpreted as the outcome of a rational economic decision-making process~\cite{harper2005economic}. Hence, by integrating the economic theory of addiction with users’ implicit behavioral data in RS~\cite{DMF}, we construct an addiction model tailored for short-video platforms. Finally, this model is trained on real-world short-video interaction data\footnote{\url{https://github.com/tsinghua-fib-lab/ShortVideo_dataset}}
 to quantify and analyze user addiction behaviors in our experiments.

The primary experimental results are illustrated in Figure~\ref{fig:intro}. Our main finding is that users’ short-video addictive behavior largely exhibits trends similar to those observed in traditional forms of addiction (e.g., substance abuse). In Figure~\ref{fig:intro}, during the initial stage (green line), users’ utility increases with longer watch time (addiction phase); after a certain point, addictive behavior reaches its peak (black dot); subsequently (red line), users begin to experience fatigue (recovery phase), eventually ceasing video consumption. Our results align with previous findings from volunteer-based studies~\cite{hamid2015blurred}.
These findings help us to build a solid simulator.

\emph{Step 2: Simulator development and application (Section~\ref{sec:simulator}).}
Based on our analysis of user addiction-aware behavior, we develop a simulator capable of learning and modeling such behavior. This simulator enables large-scale automatic evaluation and facilitates the study of addiction dynamics under different algorithms, thereby supporting the design of a more responsible and healthy web ecosystem. 
However, directly applying a large language model (LLM)-based simulator to learn the aforementioned addiction-aware objective presents a major challenge, i.e., that of \emph{personalized addiction patterns}, as different users exhibit distinct addictive behaviors, which complicate the training process.
To tackle this challenge, we introduce a training framework, \textit{AddictSim}, that integrates with a mean-to-adapted (M2A) training strategy to effectively conduct group relative policy optimization (GRPO) training~\cite{GRPO}. 
In the first stage, AddictSim predicts the average addiction-related rewards to pre-train the LLMs using GRPO. In the second stage, it refines the addiction model to estimate individualized behavioral rewards for each user and then applies these rewards across user groups, thereby improving the realism of simulated user behavior.
Experiments on two large-scale datasets verify that AddictSim outperforms other baseline training strategies. 

After constructing the simulator, it can be employed to emulate user feedback, thereby enabling automatic evaluation of algorithmic performance. Our simulations indicate that diversity-aware approaches~\cite{cpfair, xu2023p}, which enhance item category diversity, effectively redirect user attention and mitigate addictive behavior. This demonstrates the simulator’s value in realistically modeling user interactions, facilitating the testing of different recommendation algorithms, and guiding the design of healthier web platforms.

In summary, our key contributions are three-fold: 
\begin{enumerate}[label=(\arabic*),leftmargin=*,nosep]
    \item We investigate user short-video addiction behavior and formulate addiction objectives based on economic addiction theory using large-scale user data. 

    \item To build addiction-aware simulators, we propose a novel training framework, AddictSim, which incorporates an M2A strategy with GRPO training strategy. 

    \item Our experimental results demonstrate the effectiveness of AddictSim and show that, through its simulations, diversity-aware approaches can effectively mitigate addictive behaviors.
\end{enumerate}

\noindent%
To facilitate reproducibility of our work, we share our code at \url{https://github.com/XuChen0427/AddictSim}.

\vspace*{-2mm}
\section{Related Work}

\emph{Short-video recommendation.} 
Recommender systems (RS) are designed to identify and deliver items that match user preferences from a large corpus of available options~\cite{resnick1997recommender}. Traditional RSs tend to focus on domains such as e-commerce~\cite{zhao2025model} and news~\cite{wu2020mind}, where user–item interactions are usually captured by immediate actions, including exposure, clicks, and purchases. Classical user behavior modeling methods~\cite{dupret2008userbrowsing} are built on the assumption that user clicks are strongly influenced by item ranking positions, i.e., user attention decreases as the rank position goes down. With the rapid rise of short-video platforms, however, the nature of items has shifted to short videos, typically lasting less than ten minutes~\cite{GongCIKM22_ShortVideoRS, liu2019building, pan2023understanding}. Such concise content promotes rapid consumption and continuous engagement. In this setting, user behavior has evolved: instead of clicks, watch time has become the primary signal, with users often spending extended time on individual items~\cite{GongCIKM22_ShortVideoRS}. 

\emph{Short-video addiction.}  Short videos' negative impacts pose significant challenges to the sustainability of the online ecosystem~\cite{feng2024cognitive}.
Previous studies in psychology and behavioral science have shown that user behavior in short-video consumption differs fundamentally from traditional item interactions, often manifesting addictive tendencies. Such behavior is characterized by excessive and hard-to-control usage, where users become immersed in endlessly scrolling through videos to obtain instant entertainment and psychological gratification~\cite{nong2023relationship, feng2024cognitive}.
Prior research on addictive behavior has primarily relied on survey-based methods, self-reported information from volunteers, or controlled lab studies. In survey-based approaches~\cite{nong2023relationship, ye2022effects, zhang2019exploring, feng2024cognitive}, researchers typically design questionnaires to measure self-perceived levels of problematic usage and psychological dependence. Self-report studies~\cite{ye2025short} ask participants to provide retrospective accounts of their short-video usage, motivations, or feelings of loss of control. Volunteer-based experiments~\cite{chen2023effect} often recruit small groups of users to monitor their usage under specific conditions. While these methods provide valuable psychological and qualitative insights, they are limited by subjective biases (e.g., memory errors, underreporting, or social desirability effects).

\emph{Simulator learning.}
To model and observe the impact of user behavior, prior research has often employed user simulators as a  tool~\cite{Simulator_Tutorial}. Traditional simulators are typically built using statistical approaches, such as modeling user behavior by estimating action frequencies~\cite{dupret2008userbrowsing, SARDINE24}. While simple and interpretable, these methods lack generalizability and fail to capture the complexity of real-world user interactions. With the advent of LLMs and their capabilities in understanding user intent and using broad world knowledge, LLMs have emerged as a promising foundation for building user simulators~\cite{wang2024survey}. Existing studies generally adopt two approaches: (1) prompt-based simulation~\cite{zhang2024generative, RecUserSim}, where LLMs are guided by carefully designed prompts to mimic user actions; however, this approach struggles to capture user behaviors in specific domains. (2) RL–based simulation, where LLMs are fine-tuned on log data by treating user feedback as rewards (e.g., using PPO)~\cite{liu2025recoworldbuildingsimulatedenvironments}, and in some cases further incorporating user profiles to personalize the simulator according to user diverse behaviors~\cite{wang2025know}. These approaches largely overlook the modeling of personalized addictive behavior, which is critical in short-video recommendation scenarios.

\section{Preliminaries}
\label{sec:formulation}

\emph{Short-video recommendation.} 
First, we study the session-level watch time prediction task in short-video recommendation. Unlike traditional recommendation domains, short-video platforms rely on watch time as the primary signal of user engagement. Therefore, accurately predicting watch time is important for optimizing recommendation strategies and enhancing user experience~\cite{lin2024conditional, sun2024cread}.

\begin{figure*}
    \centering
    \includegraphics[width=0.85\linewidth]{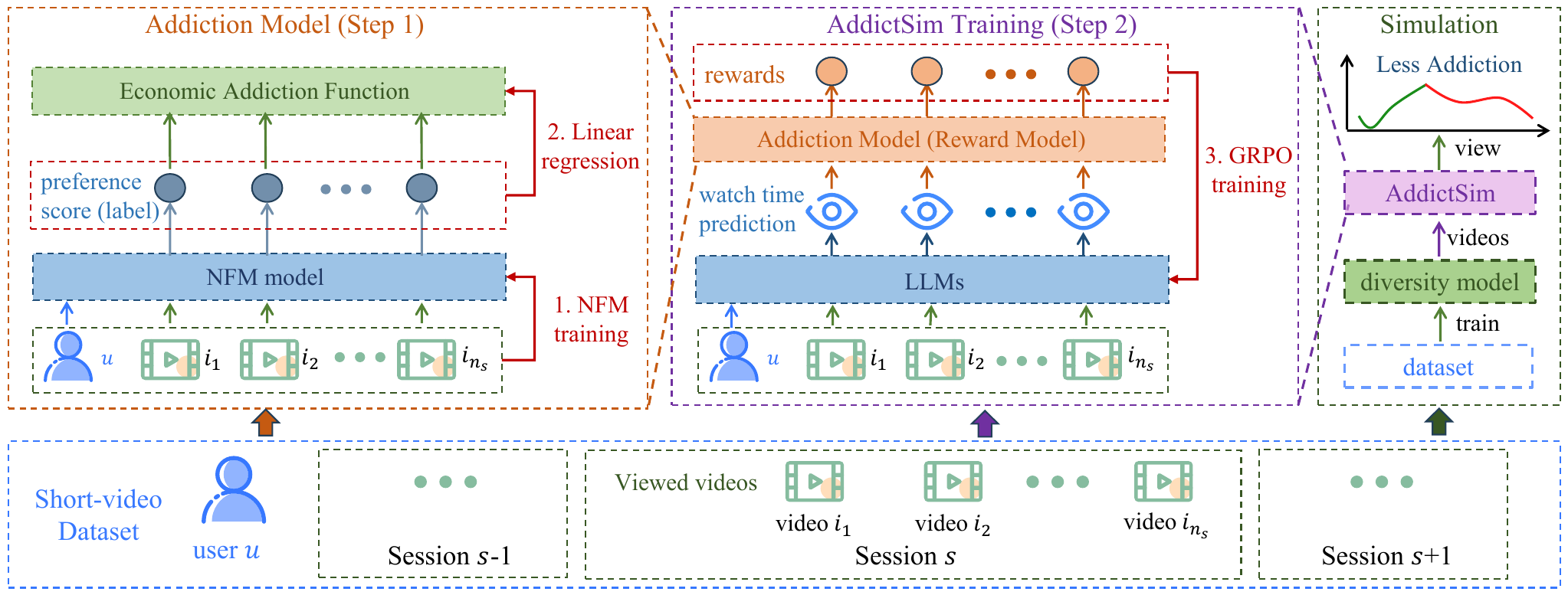}
    \caption{The framework of our study consists of two main steps. Step 1: construct an addiction model. Step 2: train a simulator, AddictSim, using the addiction model as the reward function, and apply AddictSim to run simulations aimed at mitigating addictive behavior.}
    \label{fig:framework}
\end{figure*}

Let $\mathcal{U}$ denote the set of users and $\mathcal{I}$ denote the set of candidate videos. 
For a given user $u \in \mathcal{U}$, we consider their interaction history organized into sessions, each representing a continuous watching period (\eg one hour).
A session $s$ consists of a sequence of videos watched consecutively:
\begin{equation}
I_{u,s} = \{ i_1, i_2, \dots, i_{n_s} \}, ~i_k\in \mathcal{I},
\end{equation}
where $n_s$ is the number of videos consumed in session $s\in S_u$, $S_u$ represents the set of all watching sessions of user $u$. 
For each video $i_k \in I_{u,s}$, the dataset records two types of user feedback:
\begin{itemize}[leftmargin=*]
    \item the continuous feedback, i.e., the watch time $t_{k}$;
    \item the discrete feedback: click action, denoted as
    $
    c_{k}.
    $
\end{itemize}

\noindent%
The goal of short-video recommendation is to predict both the watch time of user $u$ for a candidate video $i$ within session $s$ and the preference score for that video $i_k$.

\emph{Background: addiction models in economics.}
In economics, a rational user will maximize their intertemporal utility $u(\cdot)$ over for each period  $t=1,2,\cdots, T$:
\begin{equation}
    \max U =\sum_{t=0}^{\infty} \phi^t u(C_t, S_t),
\end{equation}
where $C_t$ denotes the consumption of an addictive good (e.g., short videos) at period $t$, and $S_t$ represents the ``addictive stock'' that accumulates from past consumption. The stock evolves: 
\begin{equation} 
    S_{t+1} = (1 - \delta) S_t + C_t, 
\end{equation} 
where $\delta \in (0,1)$ is the depreciation rate that measures how fast the addictive effect fades.
This framework highlights two key aspects: 
\begin{enumerate}[leftmargin=*]
    \item \textbf{Reinforcement}, where higher past consumption $S_t$ increases the marginal utility of current consumption (addiction phase). 

    \item \textbf{Tolerance and withdrawal}, where accumulated stock may also reduce baseline utility, leading to cycles of craving and recovery (recovery phase). 
\end{enumerate}
Such an economic addiction model provides a rigorous, micro-founded framework to capture reinforcement, tolerance, and withdrawal effects in user behavior. Its forward-looking structure allows us to analyze both short-term gratification and long-term welfare consequences of addictive consumption.

Next, we will use an addiction model to model the short-video addiction behavior in recommendations.

\section{Addiction Behavior Discovery}
\label{sec:addition}

We first integrate economic addiction theory with an RS model to construct an addiction behavior model tailored to short-video scenarios. Next, we use available short-video datasets to identify and quantify addictive behavior based on the proposed model.


\subsection{Short-video addiction mapping}
In the short-video scenario, we map economic addiction theory introduced in Section~\ref{sec:formulation} as follows:

\begin{itemize}[leftmargin=*]
    \item The utility $u(\cdot)$ of consuming a video corresponds to the user's preference $p$ for the current video.
    \item The ``addiction capital'' $S_t$ can be mapped as cumulative watch time before watching video $i_k$ in session $s$ is defined as:
    \begin{equation}
    T_k = \sum_{j<k} (1-\sigma)^jC_j,
    \end{equation}
    where $\sigma\in (0, 1)$ denotes the discounting factor, $C_j$ is the transformed watch time of video $i_j$, computed as $C_j = \log_{10}(1+t_j)$.
\end{itemize}

\noindent%
$T_k$ represents the user's \emph{addiction capital}. When $T_k$ is small, the user is more likely to be ``engaged'' and the current video watch time yields positive utility. When $T_k$ exceeds a threshold $\theta$, the user becomes fatigued, and additional watch time may yield negative utility. The correspondence between classical addiction theory and the short-video setting are summarized in Table~\ref{tab:addiction_mapping}. 

\begin{table}[h]
\centering
\caption{Mapping classical addiction theory to short-video watch behavior. ``Rec.'' is short for ``recommendation''.}
\label{tab:addiction_mapping}
\setlength{\tabcolsep}{3pt}
\begin{tabular}{p{0.46\linewidth}p{0.48\linewidth}}
\toprule
\textbf{Economic concept} & \textbf{Short-video Rec. analogy} \\ 
\midrule
Current utility $u(\cdot)$ & User preference score $\hat{p}$\\
Addiction capital $S_t$ & Cumulative video  watch time $T_k = \sum_{j<k} (1-\sigma)^j/t_j$ \\
Addiction phase & $T_k$ increases engagement \\
Recovery phase & $T_k$ reduces utility \\ \bottomrule
\end{tabular}
\end{table}


\subsection{Short-video addiction model}

The overall short-video addiction model is shown in Figure~\ref{fig:framework}. 
Next, we begin by estimating the preference scores as addiction utilities for recommendation, and then proceed to construct the addiction model and learn its parameters.

\emph{Preference score estimation.} To estimate the preference score, we decompose it into two parts as the estimated click-through rate (CTR) $\hat{c}_j$ and completion rate~\cite{lebreton2020study} $r_j$ of video $i_j$:
\begin{equation}
\hat{p}_j = \hat{c}_j + r_j, ~r_j=t_j/l_j,
\end{equation}
where $l_j$ is the video length of $i_j$. 
For CTR estimation value $\hat{c}_j$, we adopt a widely used recommendation model, the NFM~\cite{he2017neural}:  
\[
\hat{c}_j = m_0 + \sum_i m_ix_i + f_{\text{MLP}}(\bm{x};\theta_r),
\]
where $\bm{m}$ is the linear weight vector, $\theta_r$ denotes the MLP parameters and $\mathbf{x} \in \mathbb{R}^k$ denotes the latent embedding of user-item feature, and $f_{\text{MLP}}(\cdot)$ denotes a multilayer perceptron capturing higher-order non-linear interactions.  The parameters $\bm{m}, \theta_r$ can be obtained through the cross-entropy loss:
\begin{equation}\label{eq:entropy_loss}
    \resizebox{0.9\linewidth}{!}{$
    \mathcal{L}^r(\bm{m}, \theta_r) = \mathbb{E}_{u\in\mathcal{U},s\in S_u} \frac{1}{n_s}\sum_{j=1}^{n_s} \left[(1-c_j)\log(1-\hat{c}_j) +  c_j\log \hat{c}_j\right]$}.
\end{equation}

\emph{Addiction regression model.} To make the addiction model satisfy the reinforcement and tolerance property,   
we adopt a classic quadratic functional form~\cite{west2013addictiontheory} to model the relationship between current-period utility, current consumption, and accumulated addiction capital:
%
\begin{equation}\label{eq:addiction}
    \hat{p}_j(C_j, T_j) = a C_j - \frac{1}{2}bC_j^2 + wC_jT_j + \epsilon,
\end{equation}
where the regression parameter $a$ means the marginal benefit of initial consumption; $b$ reflects the diminishing satiation effect, where larger $b$ means a faster reduction in additional utility from further consumption; $w$ represents the addictive reinforcement effect of past consumption. Intuitively, $w$ means the strength of addiction. $\epsilon$ is the random noise variable. By estimating the model parameters, we can quantify the addiction intensity for specific user groups.

To estimate the parameters of the linear regression model, we adopt the commonly used \emph{least squares method}~\cite{bjorck1990least}, which minimizes the mean squared error between the predicted values:
\begin{equation}\label{eq:loss_addiction}
    \resizebox{0.9\linewidth}{!}{$\mathcal{L}^d(a,b,w) = \mathbb{E}_{u\in U, s\in S_u} \frac{1}{n_s}\sum_{j=1}^{n_s}\left[\hat{p}_j- \left(a C_j - \frac{1}{2}bC_j^2 + wC_jT_j\right)\right]^2$}.
\end{equation}

\subsection{Experimental analysis}

In this section, we will use two public short-video datasets to analyze user addiction behavior in real short-video platforms.

\subsubsection{Experimental settings} 
We use the following datasets:
\begin{description}[leftmargin=\parindent]
    \item[\rm\emph{THU}~\cite{THU_ShortVideo}] includes behavioral records from 10,000 volunteers, totaling 1,019,568 interactions on a real mobile short-video platform. The dataset spans a wide range of user profiles and video categories. 
    \item[\rm\emph{KuaiRec}~\cite{KuaiRec}] collects datya based on Kuaishou’s online short-video environment, where almost all 1,411 users have been exposed to all 3,327 items. It includes 45 fields describing the statistics of each item from July 5, 2020 to September 5, 2020.
\end{description}

\noindent%
Following the setting in~\cite{he2017neural}, we filter out users with fewer than 5 interactions in both datasets. For the remaining users, we split the data into training and testing sets with an 80\%/20\% ratio to evaluate addictive behaviors.  
Regarding session segmentation, we adopt the following rule: if the time interval between watching two videos exceeds 60 minutes, these videos are assigned to different sessions.  During the following experiments, we set $\sigma=0.9$. 


\subsubsection{Experimental results}\label{sec:addition_exp}

In this section, we aim to answer three research questions:

\begin{enumerate}[leftmargin=*,label=(RQ\arabic*)]
    \item Does addictive behavior emerge in users’ short-video consumption?  
    \item How does its extent compare to that observed in other recommendation scenarios?  
    \item Do different users exhibit heterogeneous patterns in their degree of short-video addiction?  
\end{enumerate}


\emph{Addiction existence (RQ1).} 
To answer RQ1, we aim to verify the presence of addictive behavior by analyzing the parameters learned by the addiction model. In our experiments, following Eq.~(\ref{eq:addiction}), we input each user’s session data and use the estimated preference scores $\hat{p}_j$ as regression labels to estimate the average addiction behavior for each dataset. 

\begin{figure}[h]
    \centering
    \subfigure[THU]
    {
        \includegraphics[width=0.47\linewidth]{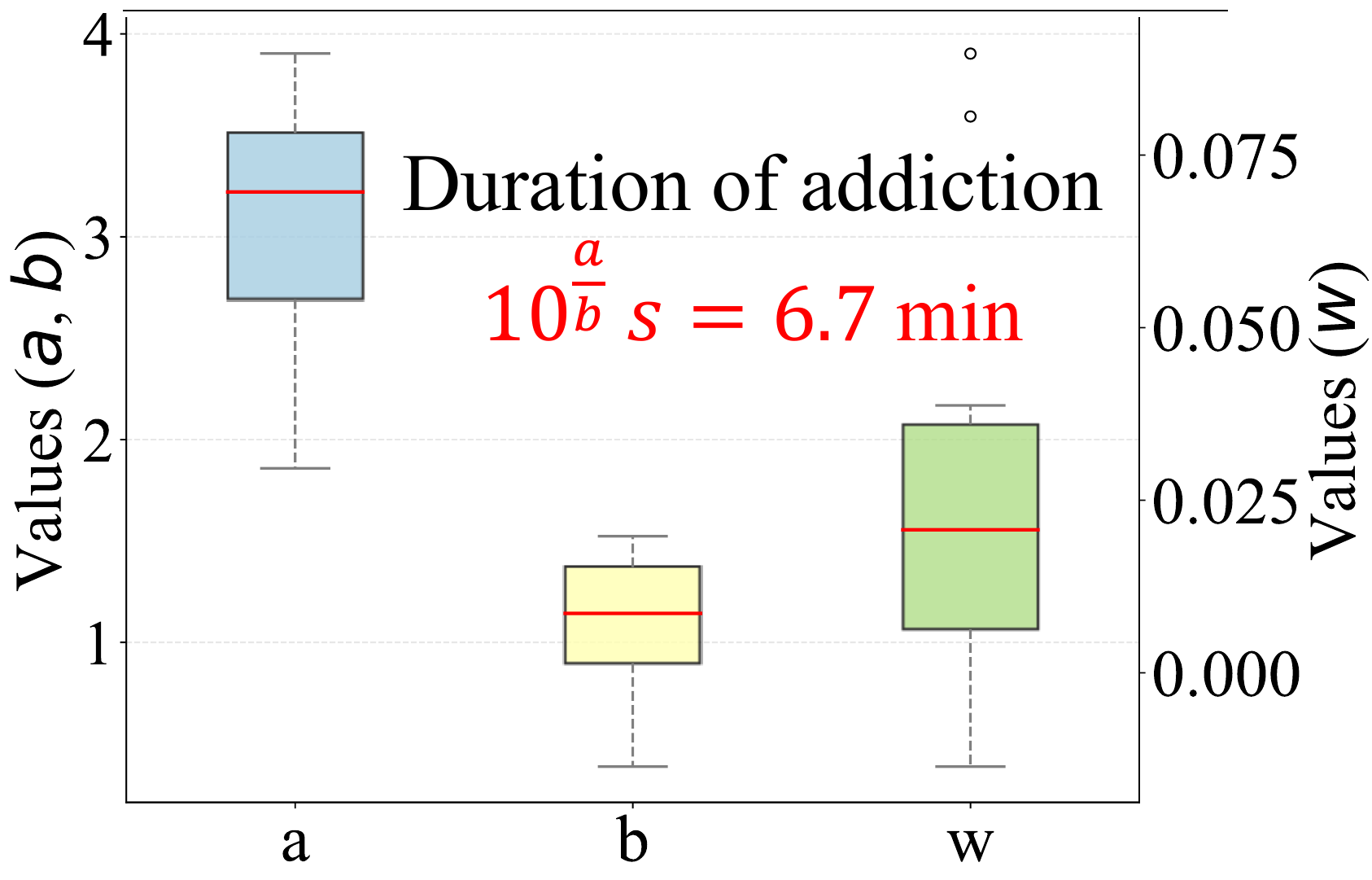}
    }
     \subfigure[KuaiRec]
    {
        \includegraphics[width=0.47\linewidth]{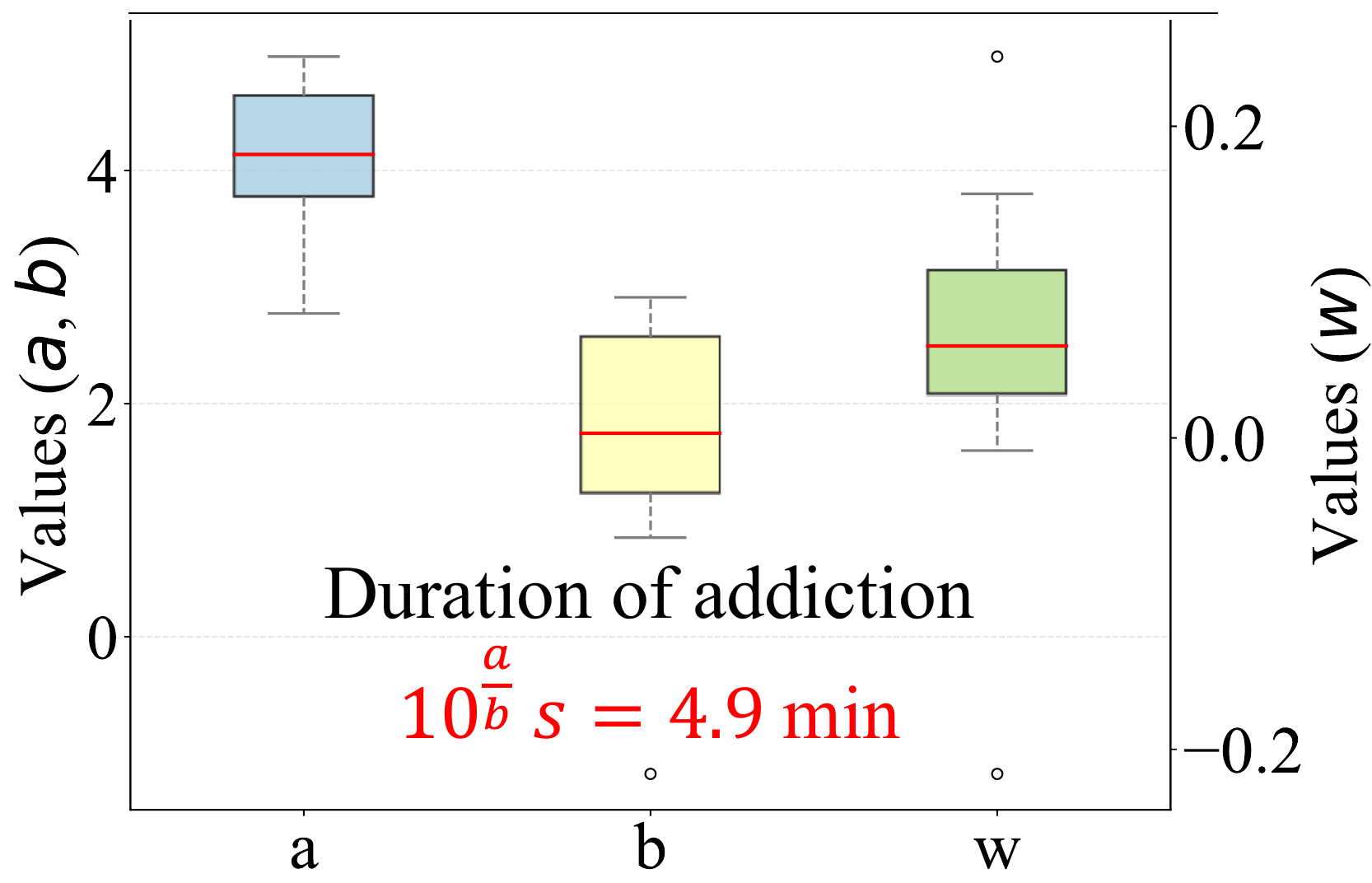}
    }
    \vspace*{-2mm}
    \caption{Average addiction model parameters $a$, $b$, and $w$ for users on the THU and KuaiRec datasets. The ratio $a/b$ represents the duration of addiction, while $w$ indicates the intensity of addiction.}
    \label{exp:addiction_exsitence}
\end{figure}

Figure~\ref{exp:addiction_exsitence} shows the boxplots of the addiction model parameters $a,b$, and $w$ for the two datasets, obtained using 10-fold cross-validation~\cite{wong2019reliable}. For each fold, we random split the dataset as $80\%-20\%$ training-test set.
The ratio $a/b$ represents the duration of addiction, while $w$ indicates the intensity of addiction.

\noindent%
From Figure~\ref{exp:addiction_exsitence}, we can clearly observe that both datasets exhibit a consistently non-zero addiction degree ($w>0$), indicating the presence of addictive behavior among users. In addition, the average duration of addiction is measured to be $6.7$ minutes for the THU dataset and $4.9$ minutes for the KuaiRec dataset, suggesting that users on these platforms tend to engage with short videos for a sustained period during each session. 
Moreover, this result is consistent with previous findings at the Arab Open University, where students were observed to watch short videos for an average of 5.63 minutes per view~\cite{hamid2015blurred}.
These findings provide empirical support for modeling user behavior with addiction-aware mechanisms.

\emph{Addiction degree (RQ2). } 
To answer RQ2, we compare users' addictive behaviors across short-video recommendation scenarios and other recommendation settings. To this end, we use the Qilin hybrid recommendation dataset~\cite{chen2025qilinmultimodalinformationretrieval}, which provides a comprehensive collection of user sessions with heterogeneous content types, including image-text notes, video notes, commercial notes, and direct answers. In addition, we consider the Ali dataset,\footnote{\url{https://tianchi.aliyun.com/dataset/56}} which samples 1,140,000 users from Taobao (a traditional e-commerce platform) over eight days of ad display and click logs, resulting in 26 million records that form the original sample skeleton. The experimental setup follows the same protocol as in previous sections.

\begin{figure}[h]
\centering
\includegraphics[width=\linewidth]{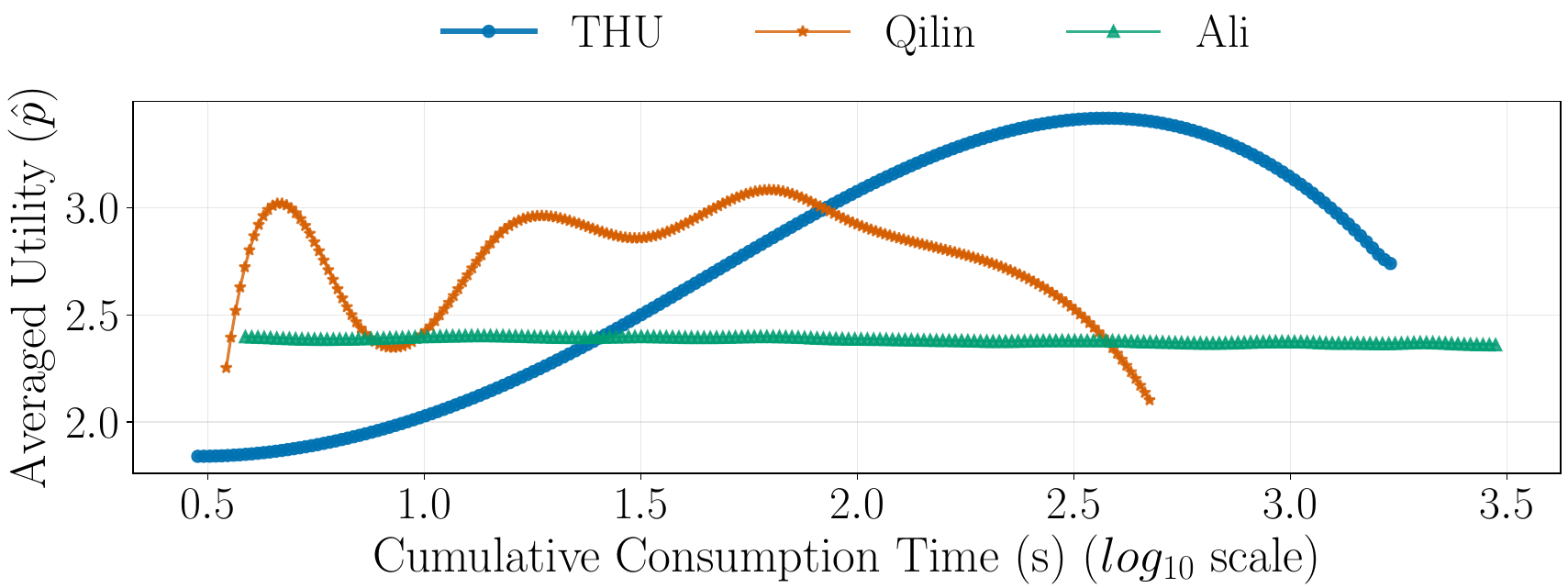}
\caption{Comparison of addictive behaviors across different recommendation scenarios, including short-video, hybrid, and traditional e-commerce settings. }
\label{exp:addiction_analysis}
\end{figure}

Figure~\ref{exp:addiction_analysis} shows the results, where the x-axis represents cumulative user watch time and the y-axis represents user utility (same settings as Figure~\ref{fig:intro}). From the figure, we can observe that in the traditional e-commerce recommendation setting (Ali dataset, green line), users exhibit almost no addictive behavior, as their marginal gains remain nearly constant over time. In contrast, the hybrid recommendation scenario (Qilin dataset, orange line) shows varying levels of addiction across different content types: for instance, users display slight addictive behavior toward text-based content in the first $30$ seconds, whereas video content induces a longer addictive period of 2--3 minutes. Finally, in the pure short-video scenario (THU dataset, blue curve), users exhibit the most pronounced addictive behavior, with the longest duration. 

\noindent%
These observations indicate that addictive behavior is a newly emerging and severe issue in short-video recommendation settings, highlighting the need to pay particular attention to this problem in such short-video scenarios.

\emph{Addiction heterogeneity (RQ3). }
To answer RQ3, we first categorized users based on their short-video viewing frequency into three groups: Low-Frequency (fewer than 100 views), Medium-Frequency (100–200 views), and High-Frequency (more than 300 views). We then examined the trends in addictive behaviors across user groups.

\begin{figure}[h]
\centering
\includegraphics[width=\linewidth]{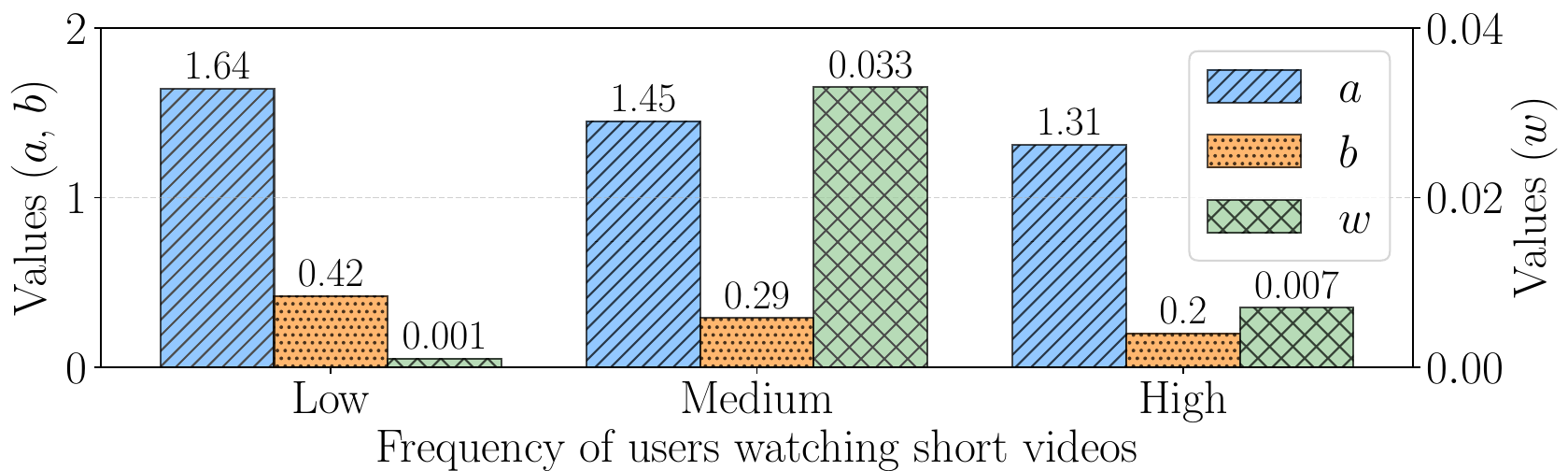}
\caption{Addictive behaviors among different user groups. }
\label{exp:frequency_users_addict}
\end{figure}

Figure~\ref{exp:frequency_users_addict} presents the results, where the x-axis represents Low-Frequency, Medium-Frequency, and High-Frequency users, and the y-axis shows the parameters 
$a,b$, and $W$ from the addiction model. From the figure, we observe that for the addiction intensity ($W$), users who rarely watch short videos exhibit minimal addictive behavior ($W$ is close to 0), whereas Medium-Frequency users show the highest addiction intensity, although their continuous addictive duration ($a/b$) is relatively short. In contrast, High-Frequency users display moderate addiction intensity, but their continuous addictive duration is relatively long ($a/b$ is larger).

Furthermore, we analyzed users’ additional profiles, as shown in Table~\ref{tab:user_addict}. Specifically, we examined the proportion of addictive users ($w>0$) across different attributes, including gender (male, female), age (18–35 as young-aged, 36–55 as middle-aged, and 55+ as old-aged), residence (urban and rural), and income level (mobile phone price <2k as low-income, 2k–4k as middle-income, and 4k+ as high-income).

\begin{table}[h]
\caption{Proportion of addictive users ($w>0$) across different user attributes, including gender, age group, place of residence, and income level.}
\label{tab:user_addict}
\setlength{\tabcolsep}{3pt}
\begin{tabular}{ccccc}
\toprule
Male   & Female & Young-Aged & Middle-Aged   & Old-Aged    \\
\cmidrule(r){1-2}
\cmidrule(r){3-5}
49.3\% & 50.7\% & 35.9\%     & 34.1\%        & 30.0\%      \\
\midrule
Urban  & Rural  & Low-Income & Middle-Income & High-Income \\
\cmidrule(r){1-2}
\cmidrule(r){3-5}
46.4\% & 53.6\% & 35.0\%     & 32.0\%        & 33.0\%     \\
\bottomrule
\end{tabular}
\end{table}

\noindent%
From Table~\ref{tab:user_addict}, we can observe that addictive behaviors vary across different user attributes. In terms of gender, the differences are minor, though female users exhibit slightly stronger addictive tendencies. Regarding age, younger users display significantly more severe addictive behaviors than older users. For residence and income level, users living in rural areas and those with lower income levels show a stronger propensity for addiction. These findings are consistent with prior research, suggesting that young and low-income users are more susceptible to addictive behaviors, and therefore represent a group that requires greater protection.

These observations indicate that users differ in both the intensity and duration of addictive behaviors, motivating the need for a simulator that can capture such heterogeneity. To this end, we propose a simulator to model user-specific addiction attributes.
\section{Simulator: AddictSim}\label{sec:simulator}

After constructing the economic addiction model in Eq.~(\ref{eq:addiction}), we use it as the \emph{reward model} to train an LLM-based simulator, termed \textbf{AddictSim}. 


\begin{algorithm}[t]
    \caption{AddictSim Training Strategy}
	\label{alg:M2A}
\begin{minipage}{0.95\linewidth}
\begin{algorithmic}[1]
\REQUIRE User set $\mathcal{U}$, user session set $S_u$, video set $\mathcal{I}$; The watching completion rate $r$ for each video;  Discounting factor $\eta, \phi$;
GRPO hyper-parameters $\epsilon, \beta, \cdots$. 
    \ENSURE The LLM simulator parameter $\theta$.
    \STATE Initialization parameter $\theta$.
    
     $// ~~ \texttt{Preference score estimation:}$
    \STATE Training RS model through minimizing the loss $\mathcal{L}^r$ in Eq.~(\ref{eq:entropy_loss})
    \FOR{$u\in\mathcal{U}, i_j\in S_u$}
        \STATE Predict the CTR score $\hat{c}_j$ using learned RS model
        \STATE Get preference score: $\hat{p}_j = \hat{c}_j + r_j$
    \ENDFOR

    $// ~~ \texttt{Stage-1 training:}$

     \FOR{$u\in\mathcal{U}, i_j\in S_u$}
    \STATE Get reward model: $\bar{a}, \bar{b}, \bar{w} = \argmin \mathcal{L}^d(a,b,w)$
    \STATE Get session-level reward: $R^1_s = $\\
    $ \sum_{j=1}^{n_s} \phi^j\left(\bar{a} C_j - \frac{1}{2}\bar{b}C_j^2 + \bar{w}C_jT_j\right)$
    \STATE Get video-level reward: $R_{*,j}^1 = \eta^{n_s-j} R^1_s$
     \ENDFOR

    \STATE We update the LLM parameters $\theta$ by optimizing the GRPO loss in Eq.~(\ref{eq:GRPO}), where the reward $R_{*,j}^1$ is employed.

     $// ~~ \texttt{Stage-2 training:}$

     \FOR{$u\in\mathcal{U}$}
    \STATE Get reward model: $a_u, b_u, w_u = $\\ 
    $\argmin \mathbb{E}_{s\in S_u} \frac{1}{n_s}\sum_{j=1}^{n_s}\left[\hat{p}_j- \left(a C_j - \frac{1}{2}bC_j^2 + wC_jT_j\right)\right]^2$
    \FOR{$s\in S_u$}
    \STATE
    \mbox{}\hspace*{-3mm}
    Reward: $R_{*,j}^2 = \eta^{n_s-j} \sum_{j=1}^{n_s} \phi^j\left(a_u C_j \!-\! \frac{1}{2}b_uC_j^2 + w_uC_jT_j\right)$
    \ENDFOR
     \ENDFOR

    \STATE We update the LLM parameters $\theta$ by optimizing the GRPO loss in Eq.~(\ref{eq:GRPO}), where the reward $R_{*,j}^2$ is employed.
    
	\end{algorithmic}
\end{minipage}    
\end{algorithm}

\subsection{Economic addiction model as reward model}\label{sec:reward_model}
In this section, we introduce the reward model, which is derived from the economic addiction model described in Section~\ref{sec:addition}. By incorporating this addiction-aware design, the LLM-based simulator can effectively learn and reproduce user addiction behaviors.

Specifically, addiction behavior is defined at the session level. For each session $s$, we define the stage-1 and stage-2 rewards as $R_s^1$ and $R_s^2$, respectively.

\subsubsection{Stage-1 Reward $R^1$}

In this stage, our objective is to train the LLM to capture the average user performance by designing an average reward, which is carried out through the following steps:

\emph{Shared addiction model.} First, we use user data to train a shared addiction model parameter:
\begin{equation}
    \bar{a}, \bar{b}, \bar{w} = \argmin \mathcal{L}^d(a,b,w),
\end{equation}
where $\mathcal{L}^d(a,b,w)$ is defined in Eq.~(\ref{eq:loss_addiction}).

\emph{Reward model.} Then, we employ the addiction model as the session-level reward function. For each session $s$, we calculate the reward $R^1_s$ based on the addiction utility, which reflects the overall satisfaction of watching within that session:
\begin{equation}
    R^1_s = \sum_{j=1}^{n_s} \phi^j\left(\bar{a} C_j - \frac{1}{2}\bar{b}C_j^2 + \bar{w}C_jT_j\right).
\end{equation}

\subsubsection{Stage-2 Reward $R^2$} 
In this stage,  our objective is to train the LLM-based simulator to capture personalized addiction behavior. 

\emph{Personalized addiction model.} First, we use each user $u$'s data to train a personalized user addiction model parameter:
\[
    a_u, b_u, w_u = \argmin \mathbb{E}_{s\in S_u} \frac{1}{n_s}\sum_{j=1}^{n_s}\left[\hat{p}_j- \left(a C_j - \frac{1}{2}bC_j^2 + wC_jT_j\right)\right]^2,
\]
where $a_u, b_u, w_u$ denote the user-specific model parameters.

\emph{Reward model.} Similar to stage-1, we construct the session-level reward based on the user’s personalized addiction model:
\begin{equation}\label{eq:reward_stage_2}
    R_s^2 = \sum_{j=1}^{n_s} \phi^j\left(a_u C_j - \frac{1}{2}b_uC_j^2 + w_uC_jT_j\right), s\in S_u.
\end{equation}

\subsection{M2A training strategy}

In Section~\ref{sec:addition_exp}, to enable an LLM-based simulator with personalization capabilities, we introduce the mean-to-adapted (M2A) training strategy with the introduced reward model. M2A first learns an average behavioral pattern and then adapts to user-specific rewards to achieve personalization. The overall workflow can be seen in Figure~\ref{fig:framework}, and the overall training procedure can be seen in Algorithm~\ref{alg:M2A}.
Both stages follow the GRPO training procedure outlined below. 

\emph{GRPO training.} Let $\pi_\theta(t\mid u,i_{<j})$ denote the LLM-based simulator parameterized by $\theta$. It takes the optional user profile and the user's least $5$ historical viewing videos $i_{<j}$ as prompt input at step $j$ of each session and outputs the predicted video watch time $t$. The detailed prompts can be seen in Appendix~\ref{app:prompt}.
And let $\pi_{\theta_{\text{old}}}$ be the behavior policy used to collect data.

Then we adapt the GRPO objective~\cite{GRPO} to fit our short-video recommendation scenarios. The objective can be written as:
\begin{equation}\label{eq:GRPO}
    \begin{aligned}
           &\mathcal{L}(\theta) =  \mathbb{E}_{u\in \mathcal{U}, s\in S_u, \{t_j\}_{j=1}^G \sim \pi_{\theta_{\text{old}}}(t\mid u,i_{<j})}\\
            &\frac{1}{G}\sum_{j=1}^G
            \{
            \text{min}\left[\widetilde{\pi}_j\hat{A}_{g,j}, \text{clip}\left(\widetilde{\pi}_j,1-\epsilon, 1+\epsilon\right)\hat{A}_{g,j}\right]-\beta\mathbb{D}_{KL}\left[\pi_{\theta}||\pi_{ref}\right]\},
    \end{aligned}
\end{equation}
where $\epsilon$ and $\beta$ are hyper-parameters, $G$ is the group sample size and
$
    \widetilde{\pi}_j = \frac{\pi(t\mid u,i_{<j})}{\pi_{\theta_{\text{old}}}(t\mid u,i_{<j})}
$ 
means the normalized output from LLM. The $\hat{A}_{g,j}$ is the advantage function calculated based on the relative rewards of the outputs inside each group $g$:
\begin{equation}\label{eq:advantage}
    \hat{A}_{g,j} = \frac{R_{g,j}^m-\text{mean}(R^m_{g,j})}{\text{std}(R_{g,j}^m)},~m=1,2, 
\end{equation}
where mean($\cdot$) and std($\cdot$) refer to computing the average and variance of group rewards $\{R_{g,j}\}_{g=1}^G$. The $R^m_{g,j}$ is defined as:
\begin{equation}\label{eq:reward_stage_1}
    R_{*,j}^m = \eta^{n_s-j} R^m_s,
\end{equation}
where $\eta$ is the discounting factor, $R^m_s$ is the session-level introduced in Section~\ref{sec:reward_model}. Note that to alleviate the sparse-reward issue, we distribute the session-level reward to each video within the session by applying a discounted reward allocation scheme. 

%

\subsection{Experimental results}

We use the two aforementioned short-video datasets, THU and KuaiRec, to verify the effectiveness of our framework M2A.

\subsubsection{Experimental settings.} We summarize the settings of the baseline and the implementation details.

\emph{Baseline strategies.} We choose the following commonly used simulator training strategies: \emph{Base Model}: The model is not trained, and directly relies on prompts to predict the user’s video-watching duration. In the main experiments, we use Qwen3~\cite{bai2023qwen} as a base model;
\emph{SFT}~\cite{openai2024gpt4technicalreport}: The model is trained in a supervised manner, where the ground-truth user watch time is used as the label;  
\emph{PPO}~\cite{PPO}: An actor-critic reinforcement learning algorithm that is widely applied during the RL fine-tuning stage of LLMs;
\emph{GRPO}~\cite{GRPO}: An improved variant of PPO that replaces the advantage function with intra-group rewards.
Note that the PPO and GRPO's reward is designed as the learned utilities of the addiction model.


\begin{table}[]
\setlength{\tabcolsep}{2pt}
\caption{Performance comparison between our simulator AddictSim and other training methods. $*$ indicates that the improvements over the best baseline (marked as underlined) are statistically significant (t-tests and $p$-value $< 0.05$).}
\label{tab:EXP:main}
\centering
\begin{tabular}{@{}lcccccccc@{}}
\toprule
 & \multicolumn{4}{c}{THU} & \multicolumn{4}{c}{KuaiRec} \\  
 \cmidrule(r){2-5}
 \cmidrule{6-9}
 & \multicolumn{2}{c}{\textbf{Session}} & \multicolumn{2}{c}{\textbf{Video}} & \multicolumn{2}{c}{\textbf{Session}} & \multicolumn{2}{c}{\textbf{Video}} \\
 \cmidrule(r){2-3} 
 \cmidrule(r){4-5}  
 \cmidrule(r){6-7} 
 \cmidrule{8-9}  
 & MAE & RMSE & MAE & RMSE & MAE & RMSE & MAE & RMSE \\
\midrule 
Base & 6.84 & 9.57 & 0.96 & 1.18 & 2.79 & 3.19 & 0.35 & 0.47 \\
STF & 3.12 & 4.64 & 0.77 & 1.01 & 2.25 & 2.60 & 0.31 & 0.42 \\
PPO & \underline{2.28} & \underline{3.70} & 0.65 & \underline{0.81} & \underline{1.96} & \underline{2.40} & \underline{0.30} & \underline{0.42} \\
GRPO & 2.33 & 3.87 & \underline{0.65} & 0.82 & 2.02 & 2.46 & 0.32 & 0.44 \\
\textbf{AddictSim} & \textbf{2.03}\rlap{$^*$} & \textbf{3.31}\rlap{$^*$} & \textbf{0.63}\rlap{$^*$} & \textbf{0.79}\rlap{$^*$} & \textbf{1.72}\rlap{$^*$} & \textbf{2.35}\rlap{$^*$} & \textbf{0.29}\rlap{$^*$} & \textbf{0.42} \\ 
\bottomrule
\end{tabular}
\end{table}


\emph{Implementation details.} As for the hyperparameters in all models, the learning rate was tuned among $[2\times10^{-5}, 2\times10^{-6}]$ and the LLM parameter $\epsilon, \beta$ were set as $0.3$ and $0.9$, respectively. The batch size is tuned among $[4,64]$, and the propagation rate is tuned among $[0.0,1.0]$. The time discounting factor $\sigma$ is tuned among $[0.1,0.9]$.
All the experiments used the Low-Rank Adaptation (LoRA)~\cite{hu2021loralowrankadaptationlarge} technique to learn the parameters.
We implement the training process using the verl~\cite{verl} framework. All experiments were conducted on a Linux platform, with four NVIDIA A800 GPUs.

\subsubsection{Evaluation metrics} Since the primary objective of the simulator in this study is to predict users' video-watching duration. Here, we test on both session-level accuracy and video-level accuracy. Follow prior work~\cite{lin2023tree} and adopt mean absolute error (MAE) and root mean squared error (RMSE) as evaluation metrics. These metrics measure the discrepancy between the simulator-predicted watch time and the actual user watch time.
We will both evaluate the session-level and video-level accuracy:

\emph{Session-level.} Since we focus on session-level addiction behavior, we evaluate the prediction accuracy as:
\begin{align}
    \text{MAE} ={}& \text{Mean}(|\textstyle\sum_{j=1}^{n_s}\hat{c}_j \!-\! \sum_{j=1}^{n_s}c_j |), \\
    \text{RMSE} ={} & \textstyle\sqrt{\text{Mean}[( \sum_{j=1}^{n_s}\hat{c}_j -\! \sum_{j=1}^{n_s} c_j )^2]}. 
\end{align}

\emph{Video-level.} For the video level, we simply predict the prediction accuracy of individual videos:
\begin{align}
    \text{MAE} = \text{Mean}\left(\left|\hat{c}_j\!-\! c_j \right|\right),~ 
    \text{RMSE} = \textstyle\sqrt{\text{Mean}[(\hat{c}_j - c_j )^2]}. 
\end{align}


\subsubsection{Effectiveness of proposed AddictSim}

Next, we present experiments to evaluate the effectiveness of the proposed simulator, AddictSim. The overall results are summarized in Table~\ref{tab:EXP:main}, which compares AddictSim with baseline training methods across two datasets, measured by MAE and RMSE at both the session- and video-level. Entries marked with $*$ indicate statistically significant improvements over the best baseline (t-test, $p < 0.05$), while underlined values denote the best-performing methods.

The results indicate that AddictSim significantly outperforms all baselines in terms of MAE and RMSE across both datasets. This confirms its effectiveness in improving watch-time prediction accuracy and highlights its advantage as a training framework for short-video scenarios.

\subsubsection{Experimental analysis}

In this section, we present experimental analyses using the THU dataset as a case study, while noting that similar patterns are observed across the other datasets. For other ablation studies on different LLM backbones and discounting factor $\sigma$, please reference Appendix~\ref{app:different_LLMs} and \ref{app:discounting}.

\begin{figure}[h]
    \centering
    \includegraphics[width=\linewidth]{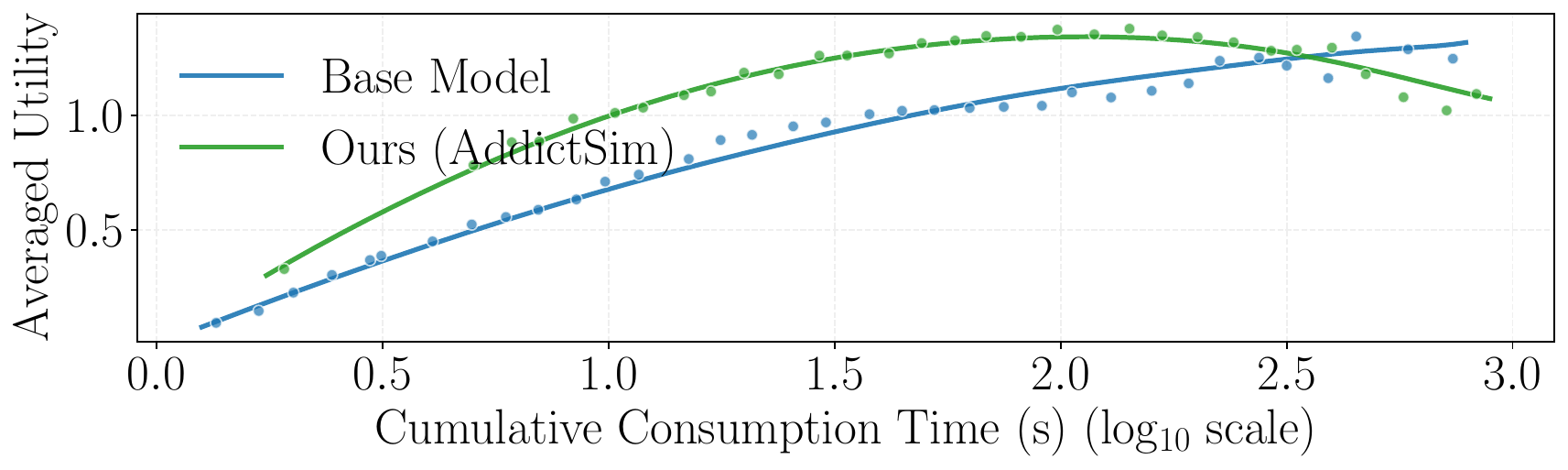}
    \caption{Addiction patterns of the simulator Base Model and AddictSim, with the x-axis representing cumulative user watch time and the y-axis representing user utility.}
    \label{exp:simulator_addict}
\end{figure}

\emph{Addiction patterns in AddictSim.} 
In this section, we aim to investigate whether the simulator AddictSim successfully captures user addiction behaviors. To this end, we plot the changes in user utility (y-axis) with respect to consumption time (x-axis) based on the simulator’s watch time predictions. Figure~\ref{exp:simulator_addict} presents the results for both the Base model (Qwen-2.5B) and AddictSim predictions.

From Figure~\ref{exp:simulator_addict}, we can observe that although the Base model prompts LLMs to learn addictive behaviors, the blue line indicates that it fails to exhibit typical addiction patterns (without recovery phases). In contrast, our simulator reproduces addiction patterns similar to those shown in Figure~\ref{fig:intro}, demonstrating that our approach effectively enables LLMs to capture user addiction behaviors.

\begin{figure}[h]
    \centering
    \includegraphics[width=0.9\linewidth]{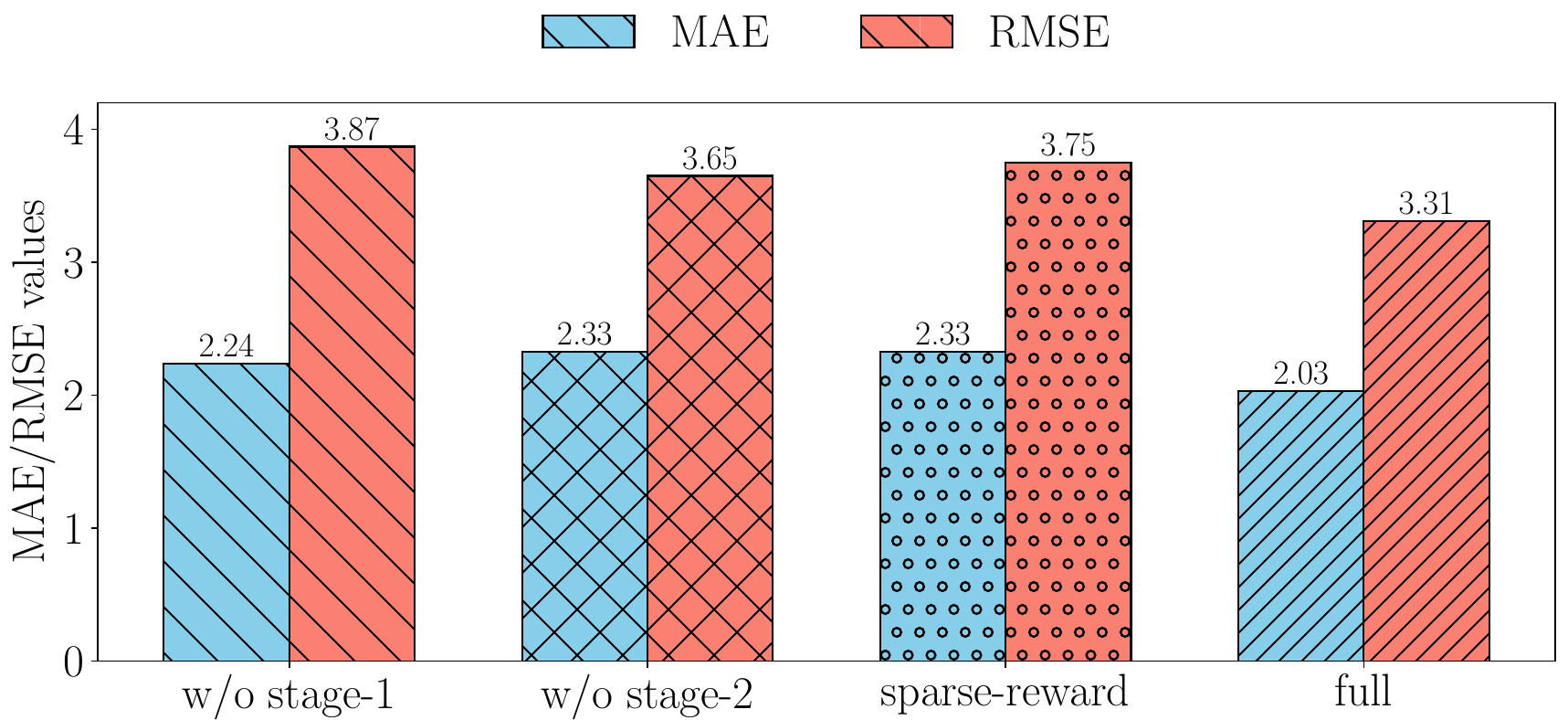}
    \caption{Ablation study at the session-level. ``w/o stage-1'' and ``w/o stage-2'' denote training without stage-1 and stage-2, respectively. ``sparse-reward'' denotes that we do not involve the discounted reward allocation scheme introduced in Section~\ref{sec:simulator}. And ``full'' indicates training with both stages. }
    \label{exp:ablation_grpo_session}
\end{figure}

\emph{Ablation study.} In this part, we aim to conduct an ablation study for AddictSim. Figure~\ref{exp:ablation_grpo_session} presents the ablation results, where ``w/o stage-1'' and ``w/o stage-2'' correspond to training without stage-1 and stage-2, respectively. ``sparse-reward'' denotes that we do not involve the discounted reward allocation scheme introduced in Section~\ref{sec:simulator}. While ``full'' indicates training with both stages. For video-level, please see the Appendix (Figure~\ref{exp:ablation_grpo_video}).

\noindent%
From the results shown in Figure~\ref{exp:ablation_grpo_session}, we observe that the ``full'' model, which includes both stage-1 and stage-2, outperforms ``w/o stage-1'' and ``w/o stage-2'', demonstrating the effectiveness of both stages. This is because stage-1 captures the average behavior of users, providing stage-2 with a solid initialization to model personalized behaviors while mitigating high variance in stage-2.
Meanwhile, we also observe that the ``full'' model outperforms the ``sparse-reward'' model, indicating that our discounted reward allocation scheme is effective and helps reduce high variance during training.

\subsection{Simulation for addiction mitigation}
\label{sec:simulation}

In this section, we aim to use our trained simulator, AddictSim, to replicate realistic patterns of user addiction behaviors. The simulator allows us to examine how users’ engagement evolves under different recommendation strategies and to assess whether these algorithms exacerbate or alleviate addictive tendencies. By simulating such interactions, we can assess the impact of algorithmic choices on user well-being and provide insights into designing RS that balance effectiveness with responsible usage.

In our simulation, we investigate whether the diversity-aware re-ranking algorithms CP-Fair~\cite{cpfair} and P-MMF~\cite{xu2023p} affect users’ addiction behaviors based on ranking results from NFM~\cite{he2017neural}. Both CPFair and P-MMF are designed to enhance provider exposure; in our setting, we define the top-level video categories as providers and aim to balance the exposure across these categories. Notably, in the THU dataset, each video is classified into a three-level categories.

For the re-ranking setup, consider a session $s$ with the original user watching sequence 
$
\{ i_1, i_2, \dots, i_{n_s} \}.
$
We can treat this sequence as a ranked list and apply a diversity-aware re-ranking algorithm that promotes lower-popularity videos to higher positions. As a result, the re-ranked sequence exposed to the user becomes:
$
\{ i'_1, i'_2, \dots, i'_{n_s} \}.
$
Then, AddictSim predicts the watching time $c_j'$ for each $j$-th video in the newly generated session sequences. Finally, we use the addiction model introduced in Section~\ref{sec:addition} to measure the addiction behaviors under different recommendation strategies.

\begin{figure}[h]
    \centering
    \includegraphics[width=\linewidth]{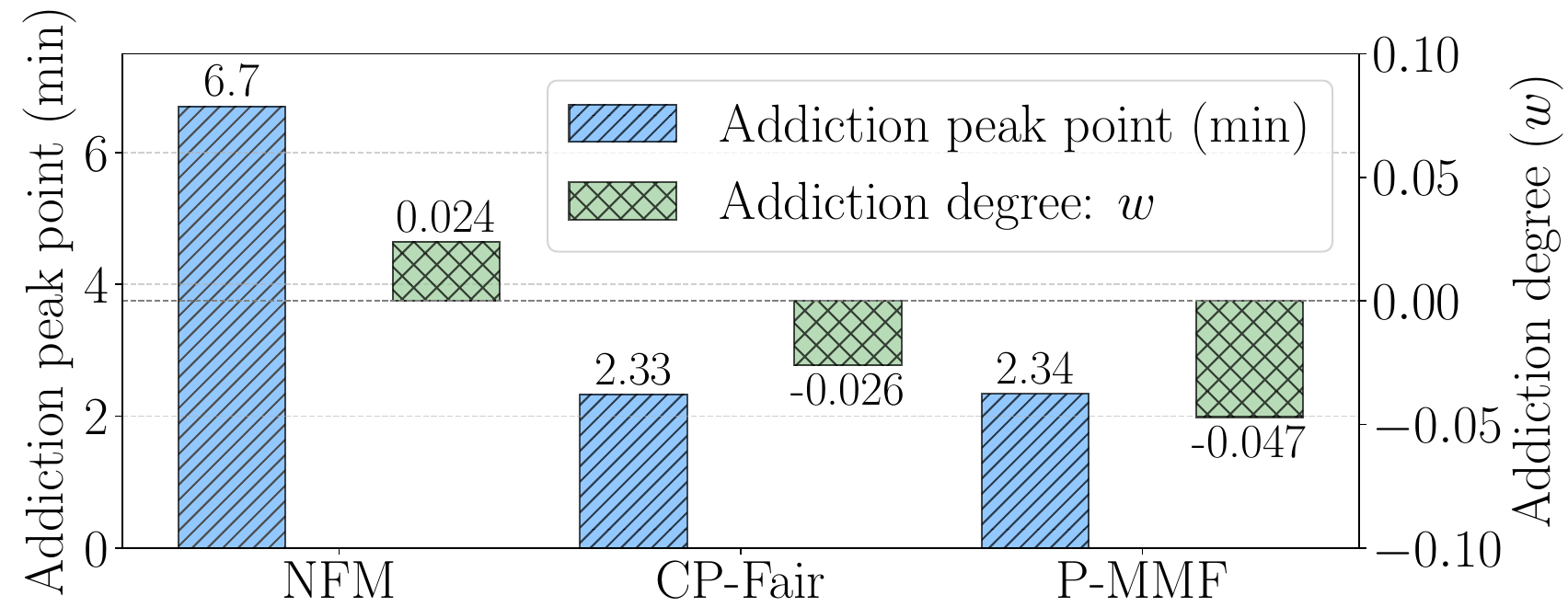}
    \caption{Addiction peak point and addiction degree for a ranking model (NFM) and two diversity-aware re-ranking models (CP-Fair, and P-MMF).}
    \label{exp:simultion}
\end{figure}

\emph{Simulation results.} Figure~\ref{exp:simultion} presents the results of simulated user addiction behaviors, measured in terms of the addiction peak point (blue bars) and addiction degree (green bars). From the results, we can clearly see that diversity-aware algorithms can substantially reduce addictive behaviors, shifting the addiction peak from 6.7 minutes to around 2.3 minutes and reducing the addiction degree $w$ from a positive to a negative value. These findings indicate that presenting users with more diverse and less-popular videos can effectively disrupt addictive patterns.

Our findings indicate that users are more susceptible to prolonged engagement when presented with similar video categories, whereas content diversification tends to mitigate such addictive behaviors. 
Such results provide practical guidance for platforms or regulatory authorities to adopt diversity-aware recommendation algorithms for teenagers who require protection. 

Furthermore, our experiments offer a useful framework for evaluating different algorithms aimed at mitigating addiction effects. We intend to further quantify the longitudinal impact of this diversity-engagement trade-off in future work.

\section{Conclusion and Discussion}

We have examined user short-video addiction behaviors through the lens of economic addiction theory, revealing severe addiction patterns using large-scale user data. We have proposed a novel simulator, AddictSim, which integrates M2A with GRPO training strategies. Extensive experimental results validate the effectiveness of our simulator. Moreover, we provide a practical example demonstrating how simulating user behavior can help identify diversity-aware approaches that effectively mitigate addictive behaviors.
Our work offers valuable practical guidance for platforms and regulatory authorities to assess addiction problems in various scenarios and to foster healthier web ecosystems.

While this paper primarily addresses anti-addiction problems, in an industrial context, the same underlying addictive mechanisms can be leveraged to optimize platform growth. For adult users, the focus may shift from total cessation of use to engagement management, particularly in mitigating the negative effects of the Recovery Phase. Our empirical findings suggest that content similarity is a primary driver of user fatigue; by identifying these saturation points, AddictSim can help design scheduling algorithms that maintain long-term retention without reaching the threshold of burnout.



\begin{acks}
    This work was funded by the National Key R\&D Program of China (2023YFA1008704), the National Natural Science Foundation of China (No. 62472426, 62572451). This research was also (partially) supported by the Dutch Research Council (NWO), under project numbers 024.004.022, NWA.1389.20.\-183, and KICH3.LTP.20.006, and the European Union under grant agreements No. 101070212 (FINDHR) and No. 101201510 (UNITE).
    All content represents the opinion of the authors, which is not necessarily shared or endorsed by their respective employers and/or sponsors.
\end{acks}

\clearpage
\bibliographystyle{ACM-Reference-Format}
\balance
\bibliography{references}

\clearpage
\appendix
\section*{Appendix}

\section{Prompt Template Used}\label{app:prompt}
We design the LLM prompt as follows. It takes the user’s viewing history and profile as input, and predicts the watch time of the current video: 

\begin{promptbox}
You are a short video user behavior prediction model. Your only goal is to predict the user's consumption for the current video.
Only output decimal Arabic numbers (with decimal points allowed), no Chinese numbers or any text.

User data (JSON):
\begin{lstlisting}[language=json]
{
  "viewing_history": [
    {
      "video_id": "33286",
      "completion_rate": 0.17,
      "categories": [" car", " say car review", " Automobile Testing"],
      "duration": 197.6,
      "watch_time": 33.0,
      "consumption_reward": 1.531
    },
    {...}]
    "current_video": {
    "video_id": "12575",
    "categories": ["history"],
    "duration": 72.4,
    "author_fans": 59478
  },
  "addiction_stock": 3.17
}
\end{lstlisting}

Please output the following field:
consumption reward: Consumption calculated as $C_t = log_{10}(1 +$ predicted watch time) (numeric value).

Strict requirements:

1) Only output the JSON below, no explanations or code blocks.

2) The JSON must be immediately followed by a line <END> as the end marker.

3) Do not output placeholders or text like "value", "NaN", "null", "None".

4) You are predicting the user's consumption Ct for the current video, not the watch time.

Correct example (example only, please output real prediction for current sample):
\begin{lstlisting}[language=json]
{"consumption_reward": <decimal float>}
\end{lstlisting}
<END>
\end{promptbox}




\section{Ablation Study at the Video-level}\label{app:ablation_video}
\begin{figure}[h]
    \centering
    \includegraphics[width=\linewidth]{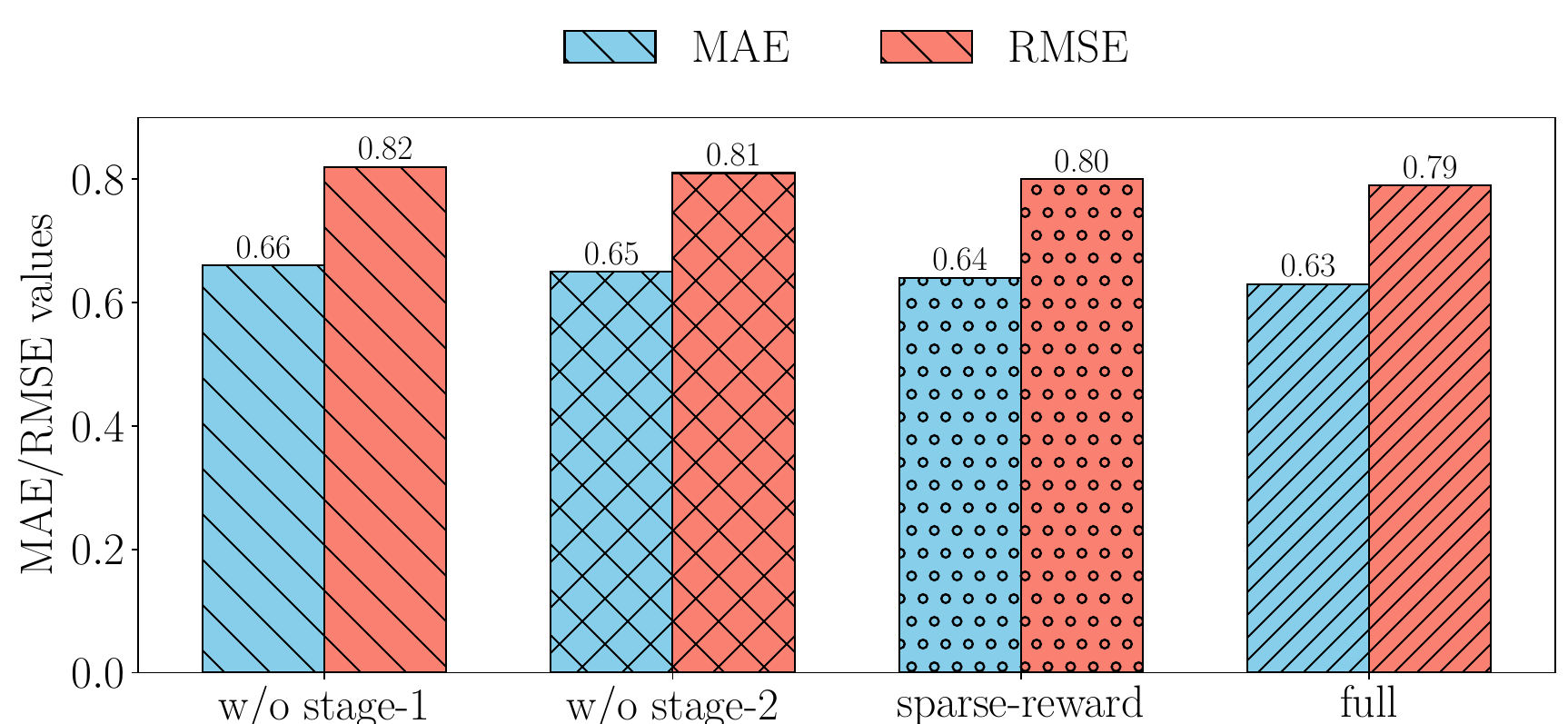}
    \caption{Ablation study on video-level. ``w/o stage-1'' and ``w/o stage-2'' denote training without stage-1 and stage-2, respectively. ``sparse-reward'' denotes that we do not involve the discounted reward allocation scheme introduced in Section~\ref{sec:simulator}. And ``full'' indicates training with both stages. }
    \label{exp:ablation_grpo_video}
\end{figure}

In this section, we present the video-level ablation study for AddictSim shown in Figure~\ref{exp:ablation_grpo_video}. The metrics are measured as the video-level MAE and RMSE. 

From the Figure~\ref{exp:ablation_grpo_video}, we can also observe that  the experimental results consistently highlight the superiority of the 'full' model; by integrating both Stage-1 and Stage-2, it achieves significant performance gains over the 'w/o Stage-1' and 'w/o Stage-2' ablations, as well as the 'sparse-reward' baseline. This performance gap underscores the critical contribution of each stage and validates that their combination creates a synergistic effect that cannot be replicated by either component in isolation.



\begin{table}[h]
\setlength{\tabcolsep}{3pt}
\centering
\caption{Session-level performance comparison with different LLM backbones on THU Dataset. $*$ indicates that the improvements over the best baseline (marked as underlined) are statistically significant (t-tests and $p$-value $< 0.05$).}
\label{tab:multi_model_session}
\begin{tabular}{lcccccc}
\toprule
& 
\multicolumn{2}{c}{Base model} &
\multicolumn{2}{c}{GRPO} &
\multicolumn{2}{c}{Ours (AddictSim)} \\
\cmidrule(r){2-3} 
\cmidrule(r){4-5} 
\cmidrule{6-7} 
 & MAE & RMSE & MAE & RMSE & MAE & RMSE \\
\midrule 
QWEN2.5-7B & 6.84 & 9.57 & 2.33 & 3.87 & \textbf{2.03}$^*$ & \textbf{3.31}$^*$ \\
Falcon3-7B & 3.50 & 5.52 & 2.26 & 3.69 & \textbf{2.07}$^*$ & \textbf{3.50}$^*$ \\
Llama-3 8B & 7.92 & 9.93 & 2.53 & 3.89 & \textbf{2.35}$^*$ & \textbf{3.75}$^*$ \\
Mistral 7B & 4.73 & 7.82 & 2.31 & 3.73 & \textbf{2.08}$^*$ & \textbf{3.44}$^*$ \\
\bottomrule
\end{tabular}
\end{table}

\section{Robustness across Different LLMs}\label{app:different_LLMs}

In this section, we present both video-level and session-level performance comparison across different LLM backbones on the THU dataset. To evaluate the robustness of our proposed method, we replace the backbone LLM with several alternatives, including Llama3~\cite{llama3}, Mistral~\cite{jiang2023mistral7b}, and Falcon3~\cite{Falcon3}. 

Note that, to evaluate the robustness of our method, we focus on comparing it with the well-performing and most relevant baselines, namely GRPO and the base model.

\begin{table*}[h]
\caption{Performance comparison between our simulator AddictSim and GRPO. $*$ indicates that the improvements over the best baseline (marked as underlined) are statistically significant (t-tests and $p$-value $< 0.05$).}
\label{tab:EXP:user_group}
\centering
\begin{tabular}{@{}lcccccccc@{}}
\toprule
 & \multicolumn{4}{c}{Urban} & \multicolumn{4}{c}{Rural} \\  
 \cmidrule(r){2-5}
 \cmidrule{6-9}
 & \multicolumn{2}{c}{\textbf{Session-level}} & \multicolumn{2}{c}{\textbf{Video-level}} & \multicolumn{2}{c}{\textbf{Session-level}} & \multicolumn{2}{c}{\textbf{Video-level}} \\
 \cmidrule(r){2-3} 
 \cmidrule(r){4-5}  
 \cmidrule(r){6-7} 
 \cmidrule{8-9}  
 & MAE & RMSE & MAE & RMSE & MAE & RMSE & MAE & RMSE \\
\midrule 
GRPO & 1.95 & 2.75 & 0.63 & 0.80 & 2.46 & 4.25 & 0.67 & 0.85 \\
\textbf{Ours (AddictSim)} & \textbf{1.64}\rlap{$^*$} & \textbf{2.20}\rlap{$^*$} & \textbf{0.61}\rlap{$^*$} & \textbf{0.77}\rlap{$^*$} & \textbf{2.20}\rlap{$^*$} & \textbf{4.03}\rlap{$^*$} & \textbf{0.65}\rlap{$^*$} & \textbf{0.82} \\ 
\bottomrule
\end{tabular}
\end{table*}

\subsection{Session-level results for different LLMs}

From Table~\ref{tab:multi_model_session}, we can observe that despite the architectural and training differences among these backbone models, AddictSim consistently achieves notable performance improvements in terms of MAE and RSME. This indicates that the effectiveness of AddictSim is not tied to any specific LLM architecture, but rather generalizes well across a diverse set of models. Such robustness suggests that AddictSim can reliably enhance prediction accuracy regardless of the underlying language model, making it widely applicable in different short-video recommendation scenarios.

\subsection{Video-level results for different LLMs}

Then, we will present the results for the video-level simulation results of different backbone LLMs.
Table~\ref{tab:multi_model_video} presents the video-level performance comparison across different LLM backbones on the THU dataset, while video-level results exhibit similar trends.

The results align with our earlier findings: despite variations in LLMs, our approach consistently yields significant improvements in MAE and RMSE. This demonstrates that the effectiveness of our method is not dependent on a particular LLM architecture.

\begin{table}[h]
\setlength{\tabcolsep}{3.5pt}
\centering
\caption{Video-level performance comparison with different LLM backbones on THU Dataset. $*$ indicates that the improvements over the best baseline (marked as underlined) are statistically significant (t-tests and $p$-value $< 0.05$).}
\label{tab:multi_model_video}
\begin{tabular}{lcccccc}
\toprule
& 
\multicolumn{2}{c}{Base model} &
\multicolumn{2}{c}{GRPO} &
\multicolumn{2}{c}{Ours (AddictSim)} \\
\cmidrule(r){2-3}
\cmidrule(r){4-5}
\cmidrule{6-7}
 & MAE & RMSE & MAE & RMSE & MAE & RMSE \\
\midrule 
QWEN2.5-7B & 0.96 & 1.18 & 0.65 & 0.82 & \textbf{0.63}$^*$ & \textbf{0.79}$^*$ \\
Falcon3-7B & 0.77 & 0.95 & 0.66 & 0.82 & \textbf{0.65}$^*$ & \textbf{0.82}$^*$ \\
Llama 3.1 8B & 1.11 & 1.38 & 0.76 & 0.98 & \textbf{0.68}$^*$ & \textbf{0.86}$^*$ \\
Mistral 7B & 0.72 & 0.89 & 0.65 & 0.82 & \textbf{0.64}$^*$ & \textbf{0.79}$^*$ \\
\bottomrule
\end{tabular}
\end{table}

\subsection{Discussion}

In Table~\ref{tab:multi_model_session} and~\ref{tab:multi_model_video}, our model outperforms all baselines across every LLM backbone on both MAE/RMSE. Our model’s performance varies by less than 10\% across backbones, compared to up to 55\% variation in the base models (common range in recommendation task), and the GRPO-based gains remain consistently within 8–12\%. These results demonstrate that our approach is both robust and architecture-insensitive.

\begin{figure}[h]
    \centering
    \subfigure
    {
        \includegraphics[width=0.47\linewidth]{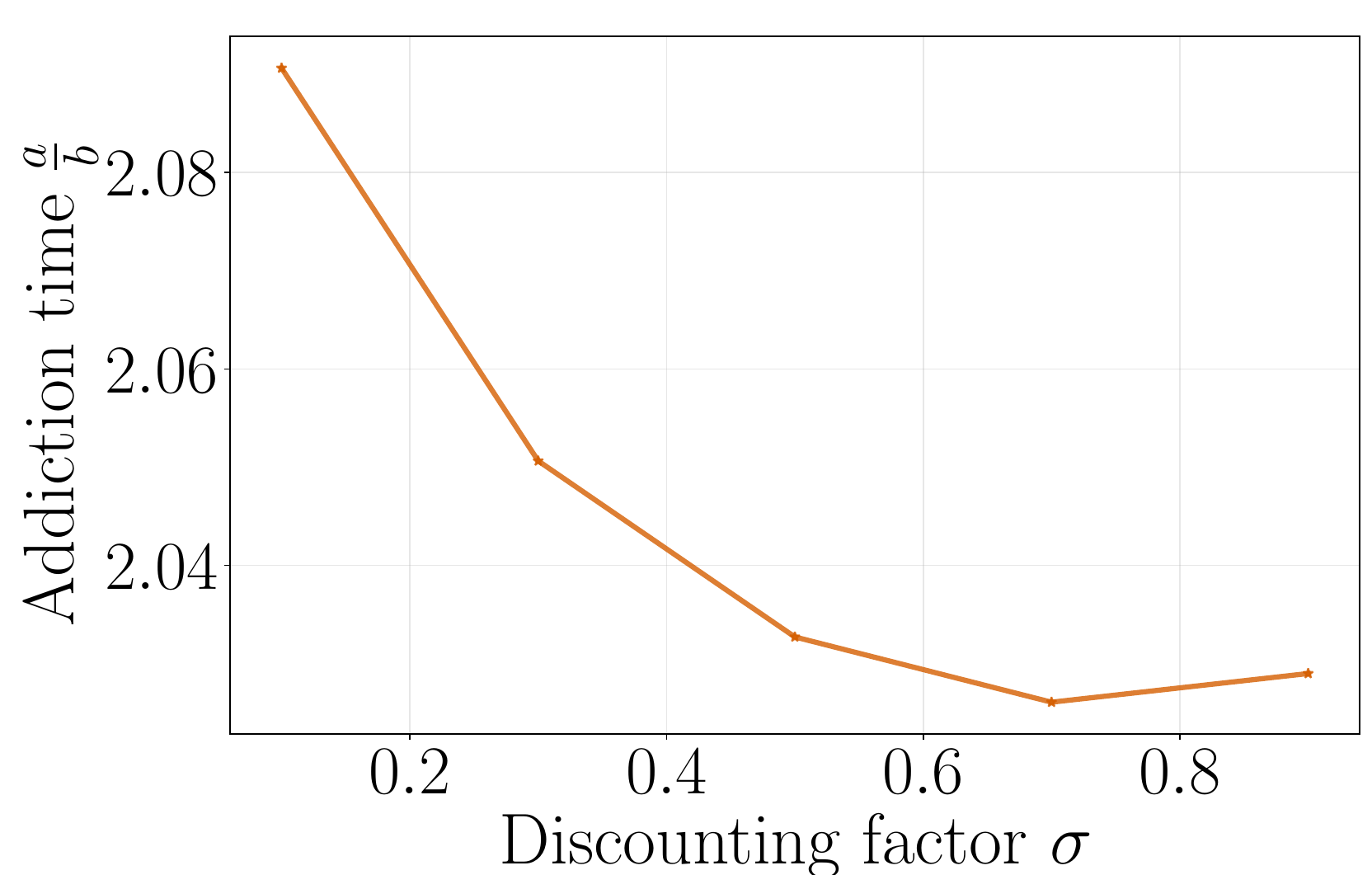}
    }
     \subfigure
    {
        \includegraphics[width=0.47\linewidth]{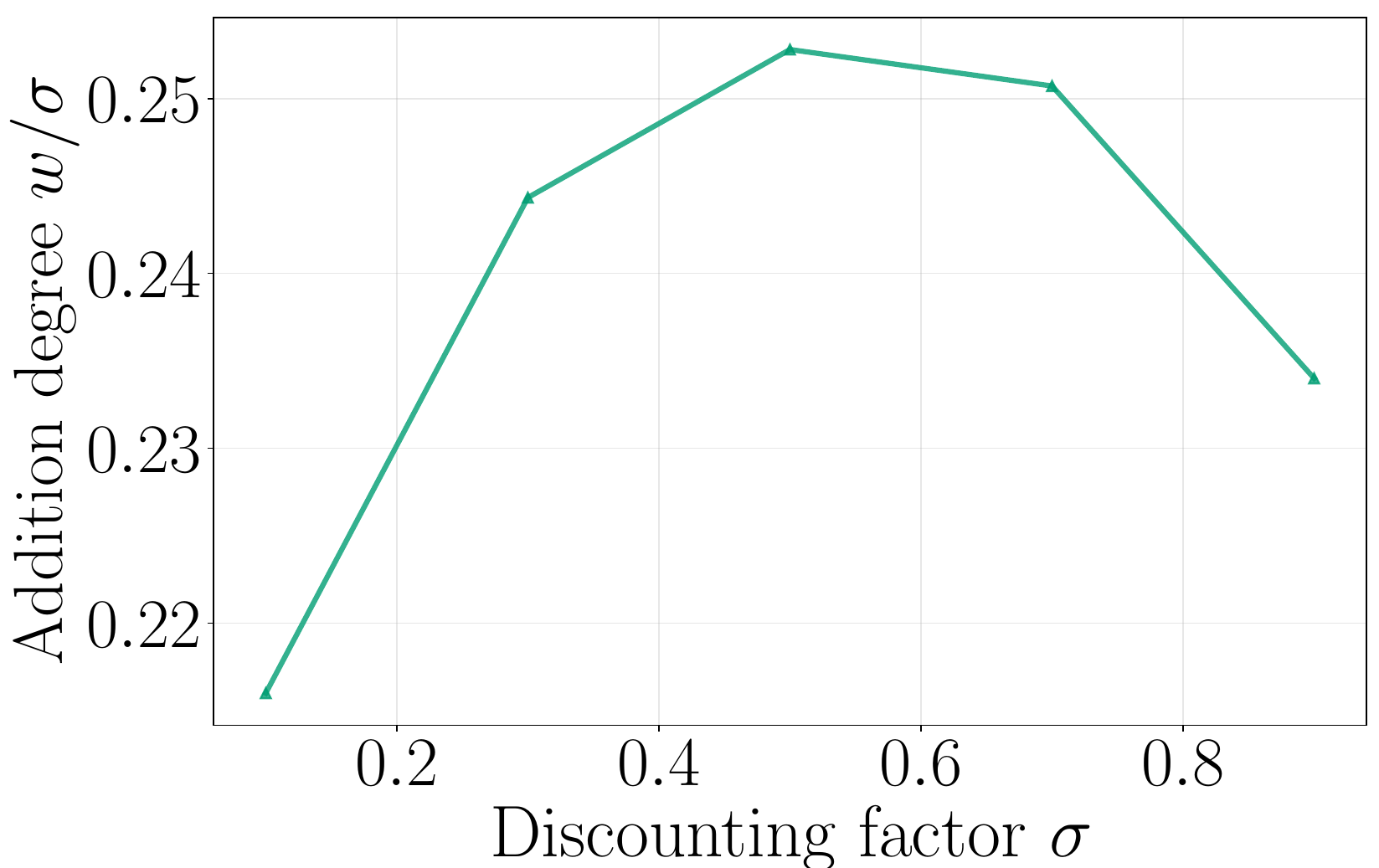}
    }
    \caption{Addiction time $a/b$ and degree $w/\sigma$ \wrt discounting factor $\sigma$. }
    \label{fig:ablation4discount}  
\end{figure}

\section{Performances on Different Users}\label{app:exp_on_different_users}

To rigorously assess the generalizability of our model across diverse user groups, we conducted a granular performance evaluation within the ``Urban'' and ``Rural'' contexts, as detailed in Section~\ref{sec:addition_exp}. This analysis allows us to determine whether the model maintains consistent efficacy across varying user groups. We present experiments to evaluate the effectiveness of the proposed AddictSim and the best-performing baseline GRPO on THU dataset. The overall results are summarized in Table~\ref{tab:EXP:user_group}. 


From the results, we can observe that our model can outperform GRPO at both session-level and video-level on different user groups. when evaluating urban and rural users separately, our model improves session-level MAE by 15.8\% and 10.5\%, respectively, over the best baseline (without considering personalized rewards). Other metrics also show consistent gains of 4–12\%.

These results further substantiate the effectiveness of our personalized reward mechanism, which is specifically designed to account for the inherent heterogeneity across diverse user groups. 

\section{Ablation Study for Discounting Factor $\sigma$}\label{app:discounting}

In this section, we investigate the influence of the discounting factor $\sigma$ on two key metrics: the addiction duration $a/b$ and the addiction degree $w/\sigma$.

As illustrated in the left panel of Figure~\ref{fig:ablation4discount}, the addiction time $a/b$ remains relatively stable across varying values of $\sigma$, exhibiting only a marginal decrease as $\sigma$ increases. In contrast, the right panel reveals a non-monotonic relationship between the discounting factor and the addiction degree $w/\sigma$. Specifically, the addiction degree initially rises with $\sigma$, reaching its peak at approximately $\sigma=0.5$, before gradually declining as $\sigma$ continues to increase.

Intuitively, $\sigma$ represents the rate at which a user "forgets" or discounts their historical consumption patterns. When $\sigma$ is low, the user is primarily driven by immediate gratification from the current video, which prevents the accumulation of deep-seated addiction. Conversely, when $\sigma$ is excessively high, the user becomes less sensitive to the immediate influence of the current video and tires of the content more rapidly. However, the influence of $\sigma$ is limited, showing the robustness of our method.


\end{document}